\documentclass[10pt, conference, compsocconf]{IEEEtran}

\usepackage{hyperref}
\usepackage{multicol}

\ifCLASSINFOpdf
  \usepackage[pdftex]{graphicx}
  \graphicspath{{../pdf/}{../jpeg/}{../png/}}
  \DeclareGraphicsExtensions{.pdf,.jpeg,.png}
\else
\fi
\usepackage{fancyhdr}
\usepackage{verbatim}
\usepackage{amsmath}
\usepackage{tikz}
\usetikzlibrary{trees}
\usepackage[english]{babel}
\usepackage{csquotes}
\usepackage{graphicx, animate}

\usepackage{media9}%

\begin{document}
\title{AI-Powered Social Bots}


\author{\IEEEauthorblockN{Terrence Adams}
\IEEEauthorblockA{tea331@g.harvard.edu}
}


%Telephone: (800) 555--1212, Fax: (888) 555--1212}


\maketitle

\begin{abstract}
This paper gives an overview of impersonation bots that generate output in one or possibly, multiple modalities.  
We also discuss rapidly advancing areas of machine learning and artificial intelligence that could lead 
to frighteningly powerful new multi-modal social bots.  Our main conclusion is that most commonly known bots are 
one dimensional (i.e., chatterbot), and far from deceiving serious interrogators.  
However, using recent advances in machine learning, it is possible to unleash 
incredibly powerful, human-like armies of social bots, in potentially well coordinated campaigns 
of deception and influence.  
\end{abstract}

\begin{IEEEkeywords}
bot, social bot, twitter bot, chatterbot, botnet, machine learning, artificial intelligence, Turing test
\end{IEEEkeywords}

% \fi
\IEEEpeerreviewmaketitle

\section{Introduction}
The term bot is a shortening of robot and refers to a software agent, or software robot device. 
It is gaining more attention very recently due to the prominence and influence 
of social media networks, as well as the ability to influence large groups through 
coordinated information operations.  The potential of deploying large influence 
campaigns has already been realized, most notably during the 2016 U.S. election 
or the preceding Bexit vote.  However, the influence of political (twitter) bots 
has been observed for years in several elections globally, including Turkey, Mexico and India.  
During the 2014 elections in India, Narendra Modi reached nearly 4 million twitter followers.  
However, when {\it Time} started monitoring Twitter for its Person of the Year award, 
the local media soon spotted a pattern.  Thousands of Modi's followers were tweeting 
the same phrase: ``I think Narendra Modi should be \#TIMEPOY'' at regular intervals, 
24 hours a day - while a rival army of bots tweeted the opposite \cite{Wool14}.

While social bots and specifically Twitter bots have demonstrated influence 
in political elections, and at the same time, in other areas (i.e., marketing), 
these bots are not known to show the sophistication that is quickly becoming 
a possibility.  As social media has grown, pushing out the impact of traditional 
print journalism, there has been tremendous growth and effort on new technological 
advances related to machine learning.  This change has been fueled by the 
availability of commodity-based distributed computing, a trend toward open source 
software development, and collection/curation of large-scale datasets.  This has enabled 
major advances in machine learning where just about any competition 
in human language technologies (speech, language, visual) is dominated 
by ``black box'' deep neural network architectures.  

\subsection*{Why now?}
Deep neural networks 
are not new to machine learning.  Frank Rosenblatt's original publication 
on the perceptron appeared in 1958 \cite{Rose58}.  
The first work detailing a learning algorithm for supervised deep feedforward 
multilayer perceptrons was developed by Ivakhnenko and Lapa in 1965 \cite{Ivak65}.  
Other deep networks date back to 1971 \cite{Ivak71}, 1979 \cite{Fuku79} and 1989 \cite{LeCu89}.  
An effective backpropagation apparently first appeared in 1970 \cite{Linn70}.  
Recurrent neural networks have also been studied such as the Boltzmann machine, 
and in 1997, the Long Short-Term Memory (LSTM) network of Hochreiter and Schmidhuber \cite{Hoch97}.  
LSTM is the basis for many current state of the art machine learning 
algorithms.  What changed the equation is not that deep neural networks 
didn't exist or weren't studied.  Instead, computing (e.g., GPGPU) and the creation 
of large-scale labeled datasets could be used to effectively train 
deep neural networks and learn a rich non-linear mapping 
from raw data (e.g., 224x224 images) to a low-dimensional representation and output layer 
(e.g., cat/not cat).



\section{A Brief History of Bots}
According to \textit{Botnets: The Killer Web Applications} \cite{Schi11}, 
``bots were originally developed as a virtual individual that could sit on an IRC channel 
and do things for its owner while the owner was busy elsewhere''.  
The first IRC bot called GM was developed in 1989, as a benevolent bot that would play 
a game of Hunt the Wumpus with IRC users. 

It was a decade later that the first botnets became known.  
A botnet is a potentially large collection of bots that communicate with each other 
or communicate with a single botmaster.  
Pretty Park was first seen in March 1999, and introduced many of the features 
that were used for more than a decade after being introduced.  It had 
the ability to report computer specifications, search email addresses, 
retrieve passwords, update its functionality, transfer files, redirect traffic, 
perform DoS attacks and communicate with an IRC server. 
SubSeven (May 1999) was the first remote controlled malware.  
The SubSeven trojan created a backdoor on the victim machine (zombie) 
by running the SubSeven server.  IRC remote control started in a later version, 
when the SubSeven server was able to receive commands via IRC.  
Many more botnets have been deployed over the years, including 
GTBot, SDBot, SpyBot, AgoBot, Rbot and Polybot.  

Botnets have continued to evolve to avoid detection, now using alternate 
communication channels, http protocol (web pages or email sites that serve up commands), 
or peer-to-peer (P2P) protocols.  They can modify DNS records and incorporate 
increasingly sophisticated techniques for command and control (C\&C).  
Botnets exhibit new forms of malicious activity, including ransomware, instant message spam, 
blog spam and distribution of disinformation and fake news. 

\section{Types of Bots}
The website botnerds.com \cite{botnerds} lists several types of bots, broken down into good and bad bots:
\begin{multicols}{2}
\begin{itemize}
% \item [\textbf{Good bots}]
\item Chatbots 
\item Crawlers 
\item Transactional bots 
\item Informational bots 
\item Etertainment bots 
% \item [\textbf{Bad bots}]
\item Hackers 
\item Spammers 
\item Scrapers 
\item Impersonators 
\end{itemize}
\end{multicols}
Chatbots or chatterbots are bots that can communicate, generally through text messages, 
with humans.  It is the type of bot typically associated with two of the best known 
Turing tests (Loebner Prize, University of Reading competition).  Possibly, the earliest 
chatbot was ELIZA which was created by Joseph Weizenbaum of the MIT Artificial Intelligence Lab 
from 1964 - 1966. 
Other well known chatbots include Cleverbot and Tay. 

Intelligent personal assistants such as Siri and Alexa act in a manner similar 
to a chatbot, but are typically much more sophisticated, communicate through audio/speech, 
and are presented as a service to users.  These digital personal assistants 
may be considered the closest broadly available artificial intelligence (AI) 
technology for providing 
human-like or human-friendly information delivery.  There continues 
to be a strong effort to teach these assistants personal skills 
(e.g., whispers, pauses, emotion \cite{Pere17}). 
% \footnote{https://techcrunch.com/2017/04/28/alexa-learns-to-talk-like-a-human-with-whispers-pauses-emotion/}
IBM Watson was originally developed as a distributed question-and-answering system, but more recently 
is being tailored to act as a chatterbot for children's toys.  Other chatterbots such as 
Eugene Goostman saw ``15 minutes of fame'' for winning the 2014 University of Reading competition.  
However, this bot, which acted the part of a 13-year old Ukranian boy, was quickly dismissed 
when exposed to the public through an Amazon Web Service.  Figure ~\ref{fig:EGbot} 
shows conversations conducted by Goostman.  
Note that Figure ~\ref{fig:EGbot} was created by G. Fariello for his 2017 course 
on comparisons between natural intelligence, artificial intelligence, 
and the potential wide-ranging implications of rapidly advancing AI \cite{Gab17}.  

User friendly applications and APIs are being developed 
for integrated delivery of chatbot services \cite{API17, slackbot}.  
Also, on the bright side, a non-profit research company, 
OpenAI, was established 
for ``discovering and enacting'' a path to safe artificial intelligence 
\cite{openAI}.  Other self-proclaimed developers of friendly bots 
come from the Botwiki community \cite{botwiki, botmakers}.  Also, 
Microsoft offers an open source bot builder SDK \cite{botframework}.

\begin{figure}
\animategraphics[height=1.9in]{.2}{goostman-}{1}{6}
\caption{\cite{Gab17} Eugene Goostman bot posing as 13-year old Ukranian boy}
\label{fig:EGbot}
\end{figure}

In this paper, we are most concerned with social bots that interact 
with social media platforms, and have the ability to impersonate users, 
while deceiving real users.  As social bots adopt 
greater levels of AI, their behaviors become increasingly difficult 
to separate from actions of real human users. 

\section{Information Operations}
There is a growing concern over the ability to unleash 
strategically mapped out and well coordinated influence campaigns 
using social bots.  In particular, Facebook appears to be responding 
to this threat (in panic mode). 
On 27 April 2017, Facebook released version 1.0 of \textit{Information Operations and Facebook} \cite{Face17}.  
%\footnote{https://fbnewsroomus.files.wordpress.com/2017/04/facebook-and-information-operations-v1.pdf''.  
This guide starts by quoting founder Mark Zuckerberg (Feb 2017):
\begin{displayquote}
It is our responsibility to amplify the good effects and mitigate the bad -- to continue 
increasing diversity while strengthening our common understanding so our community can 
create the greatest positive impact on the world.
\end{displayquote}
The document continues to define several terms bandied in the media such as: 
false news, false amplifiers, disinformation, misinformation, intent, medium, amplification.  
The article lists three major features of online information operations that 
Facebook assesses have been attempted on Facebook. 
\begin{itemize}
\item Targeted Data Collection 
\item Content Creation 
\item False Amplification
\end{itemize}
The article primarily focuses on the first and third items.  It is the second item 
which has the potential to advance rapidly based on new machine learning breakthroughs, 
and we focus on this topic in later sections. 

While information operations are gaining an increased concern globally, 
it is believed by many that this is a global threat that is taken far 
too lightly.  
Also, on 27 April 2017, the US Senate Committee on Armed Services, subcommittee on cybersecurity, 
held a hearing on cyber-enabled information operations.  
Presenting at this meeting were John C. Inglis (Former Deputy Director, NSA), 
Honorable Michael D. Lumpkin (Principal at Neptune Computer Inc and former 
Acting Under Secretary of Defense for Policy), Dr. Rand Waltzman (Senior Information Scientist, RAND Corp) 
and Mr. Clint Watts (Robert A. Fox Fellow, Foreign Policy Research Institute).  
The testimony is too long to present here, but contains alarming facts demonstrating 
the extent of information operations directed against the US and other nations.  
It calls for major changes to the US effort defending against information operations, 
and the influence campaigns conducted against our citizens. 

\subsection*{Twitter Bot Challenge}
In February/March 2015, DARPA held a 4-week competition to identify 
a set of previously identified ``influence bots'' serving as ground truth 
on a specific topic within Twitter \cite{Walt17, TBC16}.  The topic was 
pro-vaccination/anti-vaccination.  The company Sentimetrix scored at the top 
when considering the metric, 
\[
FinalScore = Hits - 0.25 * Misses + Speed. 
\]
There were a total of 39 ground truth influence bots, and Sentimetrix 
had only 1 false hit.  Participant USC correctly guessed all 39 bots 
with no false hits.  There were some clear indicators of bots 
(e.g., the variation of sentiment from the bots tends
to be lower than the variation from humans).  Humans run more hot-and-cold.  
In a paper summarizing the results of the challenge,
the following types of features were shown to be of interest \cite{TBC16}:
\begin{itemize}
\item Tweet Syntax 
\item Tweet Semantics 
\item Temporal Behavior 
\item User Profile 
\item Network 
\end{itemize}
The Twitter Bot Challenge laid ground-work for future 
evaluations of bot detection in social media.  We think 
it's important to consider much more sophisticated 
bots that are active across multiple media platforms. 

\section{Deep Learning}
The scary part of this story is that there is no evidence that today's influence 
campaigns are tapping into the rapidly advancing machine learning or AI technologies. 
The amount of research into generating signals that mimic human behavior, as well as, 
complex deep neural netowrk architectures that digest, understand and continually update 
based on human interaction, is staggering. 

It was only in 2012 that Krizhevsky, Sutskever and Hinton published 
\textit{ImageNet Classification with Deep Convolutional Neural Networks} 
\cite{Kriz12} that presented a ground-breaking leap in image recognition using 
a neural network with 60 million parameters and 500,000 neurons.  
Since then, performance has continued to improve steadily, using 
increasingly complex networks, and increasingly larger datasets.  
As an example, MSRA won the ImageNet challenge in 2015 using 
a network with 154 layers.  Google admitted to utilizing a database 
of 500 million facial images to train their face recognizer for Megaface.  
More recently, Google has announced a significant leap 
in machine translation using stacked recurrent neural networks (LSTMs) 
with both bi-drectional and uni-directional layers, as well as 
residual connections between layers \cite{Wu16}. 

\subsection{Reinforcement Learning}
The deep learning improvements highlighted above 
are supervised techniques depending on huge amounts 
of labeled data (text, image, speech or video).  
However, there have been significant advances in unsupervised 
or semi-supervised machine learning.  Techniques that utilize 
reinforcement learning accept goal oriented responses, 
and learn to optimize their decisions in complex environments.  
Probably the most recent success reported along these lines, 
is the ability of a Google DeepMind AlphaGo system to beat 
a 9-dan professional without handicaps.  
Initially, the system was developed using supervised learning 
from human play, but then it was optimized using reinforcement 
learning, by being set to play against other instances 
of itself.  
Duplicating this success does not appear very easy.  
Facebook's Darkforest system has not defeated a professional human 
Go player.  Also, Dwango and the University of Tokyo developed 
DeepZenGo which has not yet demonstrated dominance over 
top human players. 

\subsection{Generative Adversarial Networks}
A new deep learning architecture known as generative adversarial networks (GANs)
\cite{Good14} has the potential to improve recognition in multiple domains, and at the same 
time deliver systems which can generate non-human output that resembles 
human-like characteristics. 
GANs are constructed from two multilayer perceptrons G and D.  
The network $G=G(z; \theta_g)$ will take input parameters and map to the data space.  
The network $D = D(x; \theta_d)$ takes parameters and input from the data space 
and outputs a single scalar.  $D$ is trained to maximize the probability 
of assigning the correct label to both training examples and samples 
from $G$.  Simultaneously, $G$ is trained to minimize  $\log{(1 - D(G(z)))}$, 
or alternatively, maximize $\log{D(G(z))}$. 

While the discriminating network $D$ is learned, the network $G$ is improved 
to produce data that resembles the initial labeled data fed into $D$, and 
its probability distribution.  
See \cite{Radf16} and \cite{Ho16} for interesting output from some trained 
generative networks.

\begin{figure}
\includegraphics[height=1.7in]{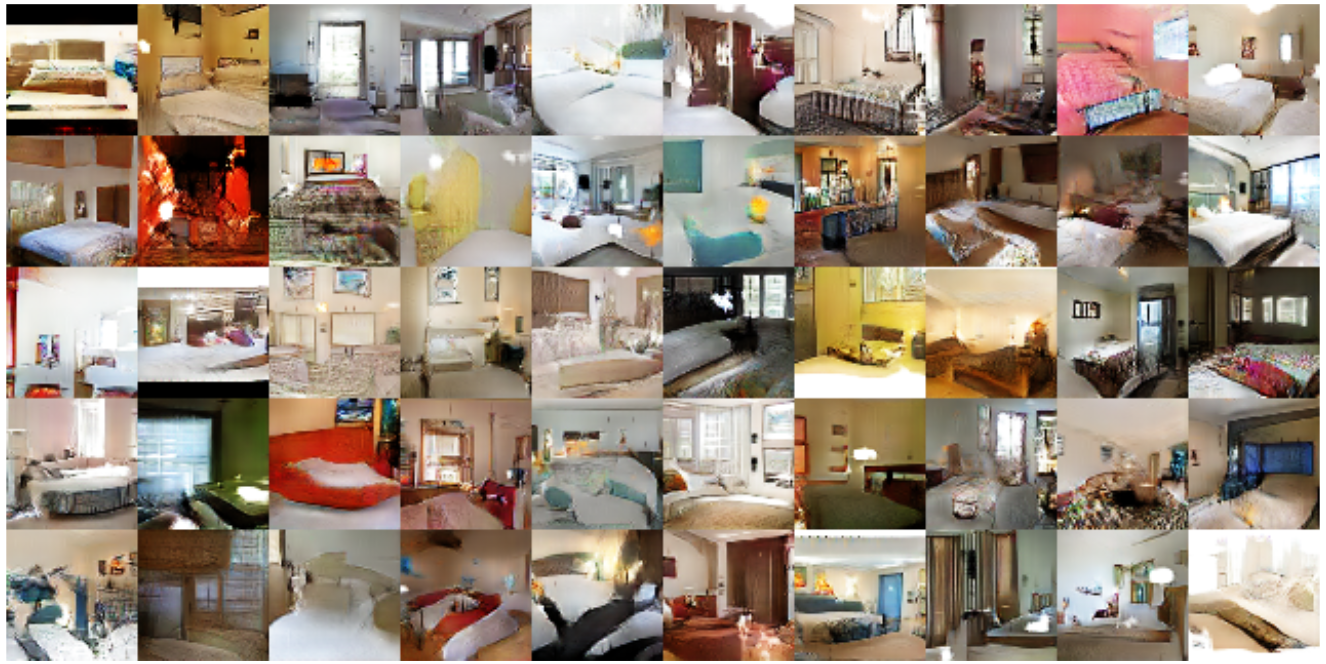}
\caption{Synthetically generated bedrooms taken from \cite{Radf16}}
\label{fig:Beds}
\end{figure}

\section{Multimodal Personas \& Scary Bots}
Deep neural networks are being applied in just about every human language technology 
to achieve state of the art for recognition.  
With companion generative networks, it will (or is already) possible 
to generate output that mimics human behavior, as well as human environments \cite{Radf16}.  
For instance, a Montreal based company Lyrebird \cite{tc042517} recently demonstrated 
the ability to take any user's voice and create a speechbot that talks 
in a similar manner to the given voice.  
Today, it is still possible for most 
human ears to detect anomalies with the bot voice that separate 
it from the real human voice, but it's uncannily close.  
With more collected data, and a larger neural network, 
it will only get better. 
Also, Google DeepMind produced a similar capability called WaveNet \cite{Wav_link}. 

It is becoming commonplace for Hollywood movies to modify facial 
gestures of actors after shooting has completed, and without 
the participation of the actual actors.  It can be done synthetically, 
and movie goers have no idea any changes were made.  
Also, there are efforts to synthetically produce digital actors/actresses 
which are amazingly close to real-life performers \cite{USC1}.

With the ability to generate text (i.e., chatterbot), speech or imagery, 
it is a matter of time before someone puts these capabilities 
together to build full walking, talking, texting social bots.  
While the replication of physical robots will be limited initially 
by the ability to manufacture material resources (e.g., metal), 
the proliferation of various social bots will only be limited 
by storage and compute power.

%\begin{figure}[!h]
%\centering
%\includegraphics[width=2.5in]{png/BED_diagram.png}
% where an .eps filename suffix will be assumed under latex, 
% and a .pdf suffix will be assumed for pdflatex; or what has been declared
% via \DeclareGraphicsExtensions.
%\caption{System Diagram}
%\label{sys_diag}
%\end{figure}



%\bibliographystyle{IEEEtran}

\end{document}